\documentclass[sigconf]{acmart}
\usepackage{balance}
\setcopyright{none}
\usepackage{balance}
\usepackage[colorinlistoftodos,prependcaption,textsize=tiny]{todonotes}
\usepackage{caption}
\usepackage{tabularx, environ}
\usepackage{titlesec}
\usepackage{float}
\usepackage{ifsym}

\begin{document}

%
\title{DeepCAT: Deep Category Representation for Query Understanding in E-commerce Search}

\author{Ali Ahmadvand}
\affiliation{Emory University\country{USA}}
\email{ali.ahmadvand@emory.edu}

\author{Surya Kallumadi}
\affiliation{\institution{The Home Depot\country{USA}}}
\email{surya@ksu.edu}

\author{Faizan Javed}
\affiliation{\institution{The Home Depot\country{USA}}}
\email{Faizan\_Javed@homedepot.com}

\author{Eugene Agichtein}
\affiliation{Emory University\country{USA}}
\email{eugene.agichtein@emory.edu}

\begin{abstract}

Mapping a search query to a set of relevant categories in the product taxonomy is a significant challenge in e-commerce search for two reasons: 1) Training data exhibits severe class imbalance problem due to biased click behavior, and 2) queries with little customer feedback (e.g., \textit{tail} queries) are not well-represented in the training set, and cause difficulties for query understanding. To address these problems, we propose a deep learning model, DeepCAT, which learns joint word-category representations to enhance the query understanding process. We believe learning category interactions helps to improve the performance of category mapping on \textit{minority} classes, \textit{tail} and \textit{torso} queries. DeepCAT contains a novel word-category representation model that trains the category representations based on word-category co-occurrences in the training set. The category representation is then leveraged to introduce a new loss function to estimate the category-category co-occurrences for refining joint word-category embeddings. To demonstrate our model's effectiveness on {\em minority} categories and {\em tail} queries, we conduct two sets of experiments. The results show that DeepCAT reaches a 10\% improvement on {\em minority} classes and a 7.1\% improvement on {\em tail} queries over a state-of-the-art label embedding model. Our findings suggest a promising direction for improving e-commerce search by semantic modeling of taxonomy hierarchies.

\end{abstract}
\maketitle              

\section{Introduction and Related Work}

\begin{figure*}
\centering
\includegraphics[origin=c,width = 450pt]{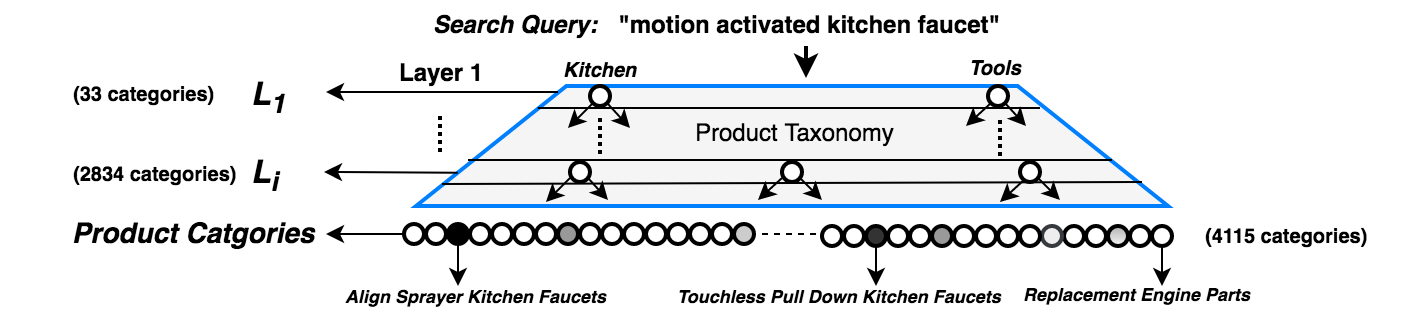}
\caption{ \small Query understanding procedure. }
\label{fig:qu}
\vspace{-2mm}
\end{figure*}

Query understanding is an essential step in developing advanced retrieval systems (e.g., e-commerce search engines) \cite{croft2010query}. In an e-commerce setting, one aspect of query understanding is achieved by mapping a query to a set of relevant product categories \cite{zhao2019dynamic}. For example, for the query `` motion activated kitchen faucet'', an e-commerce search engine should return products from relevant categories like \textit{bath, plumbing, kitchen}. These categories match the customer's intent and provide signals for downstream tasks such as retrieval and ranking. In this paper, we develop a new model for query understanding in an e-commerce search engine, depicted in Fig. \ref{fig:qu}. Fig. \ref{fig:qu} shows the query understanding procedure where a search query like ``motion activated kitchen faucet" is mapped to a set of relevant product categories in a hierarchical product taxonomy.

Query understanding is a challenging task since: 1) queries are often short, vague, and suffer from the lack of textual evidence \cite{ha2016large}, 2) queries with similar textual information with slight variations such as ``9 cu. ft. chest freezer in white'' and ``9 cu. ft. upright white freezer'' belong to different categories. However, queries with no term overlap like ``french door 32 inch. refrigerator” and ``black fridge with glass panes’’ are semantically similar and may belong to related categories, 
3) The severe data imbalance problem resulted from customer \textit{bias} towards some specific products in general or in a particular time. Also, the product categories' correlations directly impact customer click behavior, where some of them received more click rates than usual, and others get fewer click rates, and 
4) queries with low customer behavior feedback (e.g., \textit{tail} and \textit{torso}) are more challenging to classify as they have a high signal-to-noise ratio. Current neural models achieve a \textit{softer representation} with \textit{richer compositionality} of the queries compared to conventional term-based models \cite{zhang2019generic}. 

There have been numerous studies in neural models for text representation in different levels, such as characters, subwords, words \cite{Conneau:VDNN, liu2017deep, vaswani2017attention}. These models utilize distributed representation by transferring knowledge from other resources to enrich the query representation \cite{bojanowski2017enriching, kim:2014}. However, they still have difficulty properly addressing challenges (3) and (4) for query understanding. To alleviate these problems, inspired by work in information networks \cite{line:2015}, we propose a joint word-category (label) representation to provide both word and category embeddings. Then, category representations are leveraged to boost the model's efficiency on both \textit{tail} queries and the \textit{minority} classes.  

Tang et al. \cite{Tang:2015} introduce the idea of heterogeneous text network embedding to model the word and label interactions. Guoyin et al. \cite{Guoyin:2018} expand the concept to extract the relative spatial information among consecutive terms with their associated labels. Although these models leveraged the joint word-label interactions, they still lose the knowledge in label-label correlations. Extracting category (label) co-occurrence information is essential for query understanding, where product categories inherent this correlation during taxonomy formation. Product categories are not mutually exclusive and are semantically related to each other. This correlation between product categories impacts the customer click behavior, which utilizes as supervision signals in dataset generation. Thus, category co-occurrences can be used to improve the quality of \textit{minority} classes and \textit{tail} queries, where there is less customer feedback available. To this end, we consider the product categories as an undirected \textit{homogeneous} graph, where the edges represent category correlations.  

In this paper, we introduce a data-driven approach named DeepCAT for query understanding. Our model consists of a pipeline of deep learning models that utilize both word-category and category-category interactions. In summary, our contributions are: (1) proposing a novel deep learning model for joint word-category representation, and (2) introducing a new loss function to incorporate pairwise category information into the query understanding process.

\begin{figure*}[t!]
\centering
\includegraphics[origin=c,width = 490pt]{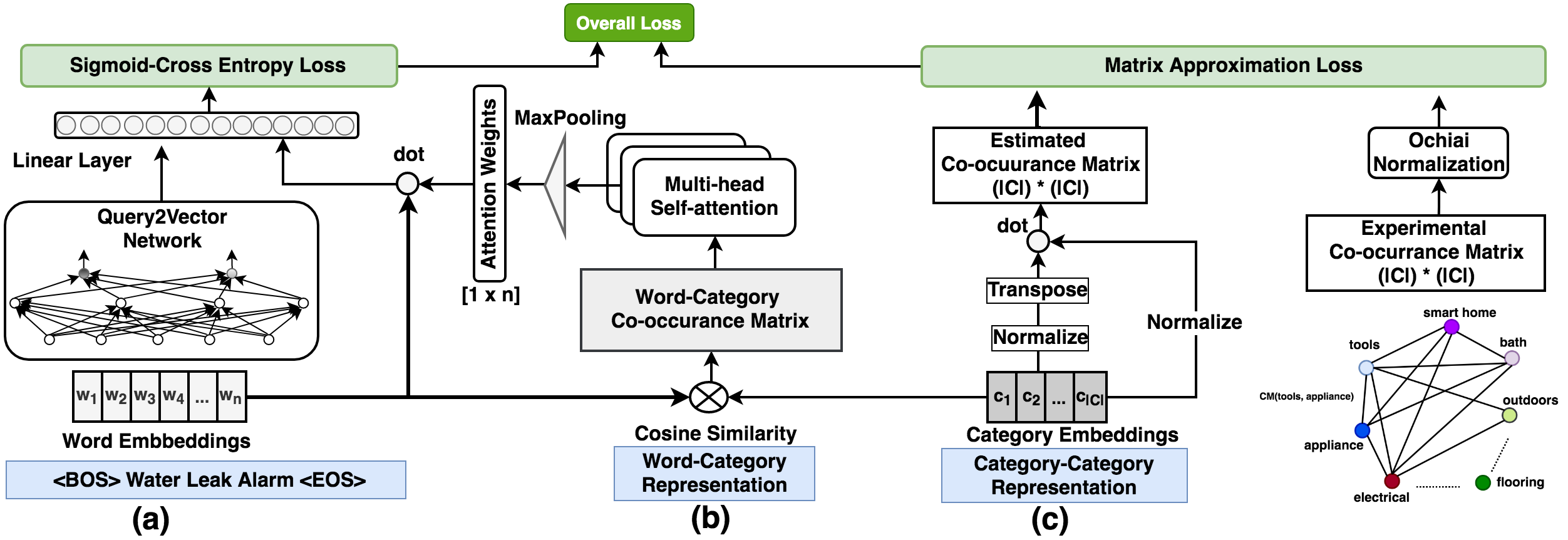}
\caption{ \small DeepCAT architecture, (a) query representation (b) word-category representation (c) category-category representation. }
\label{fig:lbl-representation}
\end{figure*}

\section{DeepCAT: Model and Implementation}

In this section, we present our DeepCAT model. First, we provide a high-level overview of the model architecture and then describe the model implementation's details. Then , we describe {\em query  representation} and {\em word-category representation}, and {\em category-category representation} models, followed by our new loss function to incorporate category co-occurrences.

\textbf{\textit{ Model Overview:} }
The DeepCAT network architecture is illustrated in Figure \ref{fig:lbl-representation}. DeepCAT consists of three main components: (a) query representation, (b) joint-word-category representation, and (c) category-category representation. Any state-of-the-art deep network could be used to develop the query representation (Query2Vector Network). We deploy a CNN-based model, which consists of convolutional layers followed by highway layers \cite{zhang2019generic}, to add more non-linearity to the model and improve the model capacity by allowing information flow in the network. Recurrent \cite{chen2019deep} or transformer \cite{vaswani2017attention} neural models could also be used as an alternative for Query2Vector network. However, due to their high latency time during inference compared to feed-forward neural models, we decided to choose the convolutional neural network-based models for query representation.

We leverage the word-category co-occurrence concepts for joint-word-category representation, which computes using a cosine similarity between query words and their associated categories. Then, a multi-head self-attention deploys to generate the contribution of each word to each specific product category. These attention scores utilize to modify the word's contribution in the query modeling. Finally, category and query representations are concatenated to create the final query representation. A sigmoid cross-entropy is deployed to compute the loss values for this multi-label problem. For category-category representation, first, we extract the experimental category co-occurrence matrix \textit{CM} from the training set. Next, it normalized using the \textit{Cosine} normalization method. In each training step, the \textit{CM} is estimated using the category representations, and the loss values are propagated through the network using matrix approximation \cite{li2018stable}. 

\textbf{\textit{ Query Representation (Query2Vector Network):}}
Suppose there is a search query dataset $ D =  \lbrace Q , C \rbrace$, where $Q$ is a set of search queries and $C$ is candidate product categories. Each query consists of a sequence of words $q = [w_1; w_2; ... \ ;w_n]$ of size $n = 10$, and is represented as $q_w^{|n| \times V}$. Also, $C$ is mapped to embedding spaces of $\mathbf{C}^{|C| \times V}$. The word and category embeddings are initialized with Word2Vec and random embedding of size $|V| = 100$, respectively. For the query representation, any complex deep learning model could be used. Our implementation of Query2Network uses a 3-layer CNN model, where it receives the word embeddings and produces the query representation. $cnn(q_w)$, goes through a {\em highway} layer \cite{zhang2019generic}. A highway layer combines a ReLU function for a non-linear projection, followed by a sigmoid function for smoothing the projection of each convolutional layer, $highway(q_w) = relu\Big(sigmoid \big(cnn \big) \Big)$.

\textbf{\textit{ Word-Category Representation:}}
To train category representation, first, in each training step, we form a word-category co-occurrence matrix. The index $(i,j)$ of this matrix indicates the co-occurrence of word $i$ and associated category $j$ of the query. To estimate this matrix during the training, we need a dot-product between word representations of query $(n \times V)$ with the category representations $(|C| \times V)$. The output is of size $(n \times |C|)$, where $n, |C|,$ and $|V|$ indicate the query length, number of categories, and embedding size, respectively. After estimating the word-category co-occurrence matrix, we need to extract each word's contribution in the query to all product categories. We deploy a self-attention mechanism with $n=10$ different heads to compute the scores. We use ten heads since we consider each query at most includes ten words. Finally, an attention matrix of size $(n \times |C|)$ creates $A_{wc} = Self\_Attention \big( l2\_norm(q_w) \odot l2\_norm(C) \big)$, where the value at $(i,j)$ represents the contribution of word $i$ to category $j$. The output goes through a max-pooling layer to form the attention weights. The attention weights multiples to the word vectors to generate the weighted word embeddings $R_{wc} = q_w \odot A_{wc}$. A multi-head self-attention mechanism applies to $q_w$. Multi-head self-attention contains several linear projections of a single scaled dot-product function that are parallelly implemented $head_i = SoftMax \Big( \frac{ q_w K^T}{\sqrt{d_k}} \Big) \mathrel{V}$. Where $\odot$ indicates a dot-product. Finally, $R_{q_{w}}$ and $R_{wc}$ go through a linear layer to form $R$, the final joint word-category representation.

\textbf{\textit{ Category-Category representation:}}
\label{cr}
A co-occurrence matrix creates on training data to model the category-category interactions. In this matrix each element $(i,j)$ represents the co-occurrence frequency between label-pair of $(c_i,c_j)$ in the training set. Finally, category-category co-occurrence matrix has the size of $|C| \times |C|$. Then, the final matrix is calculated by applying a matrix normalization. We deployed \textit{Cosine} normalization to normalize the CM, where the values on the main diagonal are one. 
Moreover, the experimental category-category CM is computed using category co-occurrences in the training set. To estimate the normalized matrix, \textit{Cosine} similarity is used between category representations.

\textbf{\textit{ Joint Word-Category Loss:}}
A sigmoid cross-entropy loss function $ \mathcal{L}_{pc}$ uses for final product category classification. Sigmoid cross-entropy applies since, in sigmoid, the loss computed for every output $s_i$ is not affected by other component values.  $\mathcal{L}_{pc} = - \sum_{c=1}^{|C|}{t_c \log \left ( {Sigmoid(s_c)}  \right ) }$. Where $s_c$ represents the predictions and $t_c$ indicates the targets.

\textbf{\textit{ Category-Category Loss: }}
The estimation error is calculated based on a matrix approximation loss \cite{li2018stable}, $\mathcal{L}_{CM} = \frac{1}{mn}\sum_{i,j \in C}log(1 + exp(\hat{CM_{ij}} \odot CM_{ij}))$.

\textbf{\textit{ The Overall Loss:}} To compute the overall loss, 
a weighted average of $\mathcal{L}_{W}$ and $\mathcal{L}_{CM}$ is computed as $\mathcal{L}_{overall} = \lambda_1 \mathcal{L}_{CM} + \lambda_2 \mathcal{L}_{W}$.

\section{Experimental Evaluation}

This section describes dataset overview, experimental design, parameter setting, metrics, baseline models, and evaluation.

\textbf{\textit{ Dataset Overview:}}
\label{dataset}
Similar to \cite{zhao2019dynamic}, we utilize customer behavior feedback (e.g., click rate) to obtain the category labels associated with each search query. We collect two weeks of search log to create both training and test sets, where the first week is used to create the training set and the second week for the test set. The training set contains more than 11M search queries. We used 25\% of the training data for validation. To generate the test set, we map queries into three different buckets using a simple query frequency. Queries with only one occurrence experimental period are considered as \textit{tail} queries; the ones between 2 and 100 impressions are counted as \textit{torso}, and the rest as \textit{head} queries. Then, to fairly evaluate the models' performance, stratified sampling \cite{babcock2003dynamic} is used to generate the test set, where we randomly select 2000 different queries from each bucket to create the test set. 

\begin{table*}
\small
\centering
\resizebox{\textwidth}{!} {\begin{tabular}{@{}l||lll|lll|lll|l@{}}
\bf Method & \multicolumn{10}{c}{\bf Leaf Nodes (Product Categories)}\\
& P@1&R@1&F1@1& P@3&R@3&F1@3&P@5&R@5&F1@5&MAP@5 \\
\hline

TF*IDF BOW & 0.783&0.259& 0.356&0.617&0.478&0.538&0.514&0.594&0.551& 0.623\\
FastText \cite{bojanowski2017enriching} & 0.856 & 0.2001& 0.324 &  0.634 &0.444&0.522&0.504&0.557&0.542&0.666\\
XML-CNN \cite{liu2017deep}& 0.875  &0.314&0.463 & 0.683&0.549& 0.609&0.568&0.666& 0.613&0.703\\
LEAM \cite{Guoyin:2018} & 0.862& 0.302&  0.447&0.676&0.531&0.595&0.566&0.651&0.606&0.697\\
\hline
DeepCAT &  \bf 0.888$^*$& \bf 0.325$^*$& \bf 0.475$^*$ &\bf 0.690&\bf 0.560$^*$&\bf 0.619$^*$&\bf 0.576&\bf 0.680$^*$&\bf 0.624$^*$ &\bf 0.717$^*$\\

\bottomrule
\end{tabular}}
\caption{\small Performances on \textit{Product Categories} with about 4200 categories. ``*'' indicates statistically significant improvements \textit{p} $<$ 0.05. }
\label{tab:overallPC}

\end{table*}

\textbf{\textit{  DeepCAT Experimental Design:}}
We designed two different experiments to evaluate DeepCAT. In the first experiment, we assess the DeepCAT capability in mapping an input query to the first level in the taxonomy hierarchies, \textit{L1}, with 33 different classes. The \textit{L1} level contains the most abstract product categories (e.g., ``appliances'', ``tools'', and ``flooring''). This experiment is mainly outlined to estimate the performance of \textit {minority} classes. The \textit {minority} classes include the categories that contain a fairly small number of samples in the training set due to customer click behavior and category overlaps or correlations. The second experiment evaluates DeepCAT on actual product categories in the last layer of taxonomy \textit{Product Categories} with 4115 distinct categories (e.g., ``replacement engine parts'', ``wood adirondack chair'', and ``window evaporative coolers'').

\textbf{\textit{Parameter Setting:}} We used an Adam optimizer with a learning rate of $\eta=0.001$, a mini-batch of size 64 for training, and embedding of size 100 for both word and category. The dropout rate of 0.5 is applied at the fully-connected and ReLU layers to prevent the model from overfitting. 

\textbf{\textit{ Evaluation Metrics:}}
Following the conventions of the search literature to evaluate DeepCAT, we reported the overall Macro- and Micro- averaged \textit{F1}, \textit{P@K, R@K, F1@K} and \textit{MAP@K} on the top-\textit{K} results. Also, query understanding is a multi-label problem; we reported precision and recall since a practical solution must cover broader possible correct categories while simultaneously keeping precision as high as possible \cite{zhao2019dynamic}. 

\textbf{\textit{Methods Compared:}}
\label{sec:comp}
We summarize the \textbf{multi-label} classification methods compared in the experimental results.

\begin{itemize}
    \item \textbf{TF-IDF + SVM:} One-Vs-Rest SVM with a linear kernel.
    \item {\bf FastText:} Text classification method by Facebook \cite{bojanowski2017enriching}. 
    \item {\bf XML-CNN:} Extreme multi-label text classification \cite{liu2017deep}.
    \item {\bf LEAM:} Word-label representation model \cite{Guoyin:2018}.
    \item {\bf DeepCAT:} The proposed word-label representation.
\end{itemize}

\subsection{Results and Discussion}

Table. \ref{tab:overallL1} and \ref{tab:overallPC} summarizes the performance of different state-of-the-art models on curated datasets described in section \ref{dataset}. The results show that DeepCAT significantly improves Macro- and Micro-average \textit{F1}, and \textit{MAP@3} by (3.6\%, 1.5\%, and 1.2\%) over LEAM, as the best model among deep networks, on \textit{L1} level. As a results, an average improvements of (6\%, 2.8\%, and 4\%) on Macro- and Micro-averaged \textit{F1}, and \textit{MAP@3} over state-of-the-art deep learning models. For \textit{product categories}, DeepCAT outperforms LEAM by (6.2\%, 4\%, 3\%, and 3\%) on \textit{F1@1, F1@3, F1@5,} and \textit{MAP@5}, respectively.

\begin{table}

\centering
\begin{tabular}{@{}l||lll}
\bf Method & \multicolumn{3}{c}{\bf First Layer (L1)}\\
& Macro-F1 & Micro-F1& MAP@3 \\
\hline

TF*IDF BOW & 0.466& 0.669 & 0.669 \\
FastText \cite{bojanowski2017enriching} &0.496 &0.686& 0.653\\
XML-CNN \cite{liu2017deep}& 0.511&0.706& 0.694\\
LEAM \cite{Guoyin:2018} &  0.521 & 0.709  &0.701\\
\hline
DeepCAT &  \bf 0.540$^*$ & \bf 0.720$^*$ &\bf 0.710$^*$\\

\bottomrule
\end{tabular}
\caption{\small Performances on \textit{L1} with 33 categories. ``*'' indicates statistically significant improvements \textit{p} $<$ 0.05. }
\label{tab:overallL1}
\vspace{-5mm}
\end{table}

\textbf{\textit{ Results on Minority Classes:}} Table. \ref{tab:overallL1} indicates that Macro-averaged \textit{F1} improves by 2\% over Micro-averaged \textit{F1}, which shows a higher impact on the minority classes. This impact is more noticeable on 8-button minority classes, where the Macro-averaged \textit{F1} for the for XML-CNN and LEAM are 0.41.01\%, 42.90\%. At the same time, this number jumps to 47.16\% for DeepCAT, which shows more than 12\% and 10\% relative improvements, respectively.

\textbf{\textit{Results on Traffic Buckets:}} Table. \ref{tab:buckets} shows the performance of the models described in section. \ref{sec:comp} across three main buckets of \textit{tail, torso,} and \textit{head}.

\begin{table}[ht]
\centering
\begin{tabular}{@{}l||l|l|l|l@{}}
 \bf Method  & \bf FastText	&\bf LEAM&	\bf XML-CNN	&\bf DeepCAT \\
\hline
Head &          0.508 &	0.563&	0.560	& \bf 0.565 \small(+0.0\%)\\
Torso &     0.584 &	0.646 &	0.648 &	\bf 0.682 (+5.3\%)\\
Tail&     0.381 &	0.337&	0.373&	\bf 0.401 (+7.1\%)\\

\hline
\end{tabular}
\caption{\small{\textit{F1@3} results on \textit{head, torso,} and \textit{tail} buckets.}}
\label{tab:buckets}
\vspace{-5mm}
\end{table}

The results show that DeepCAT significantly outperforms the other models on both \textit{tail} and \textit{torso} buckets, while it reaches competitive results to XML-CNN and LEAM on ``head'' bucket. According to higher traffic on both \textit{tail} and \textit{torso} queries, the overall performance of DeepCAT is significantly higher compared to the other models. The \textit{F1@3} is lower on \textit{head} compared to \textit{torso} queries due to a significantly higher number of correct (relevant) categories, which causes a higher \textit{P@3} and a significantly lower \textit{R@3}.

\textbf{\textit{Ablation Analysis:}}
DeepCAT is a complex model that consists of several components. We performed a comprehensive ablation study to evaluate each component's impact on the overall performance of DeepCAT. Table. \ref{tab:ablation} reports the contribution of each component on performance. The results illustrate that utilizing the category representation describe in section. \ref{cr} provides a (5.1\%, 3.2\%) improvement on Macro- and Micro-averaged \textit{F1}, respectively. Moreover, using $\mathcal{L}_{CM}$ improves Macro-averaged \textit{F1} by (2.8\%, 1.3\%), respectively.

\begin{table}[ht]
\small
\centering
\begin{tabular}{@{}l||ll@{}}
\multicolumn{1}{c||}{\bf Method}  & \multicolumn{1}{c} {\bf Macro-F1} & \multicolumn{1}{c}{\bf Micro-F1} \\
\hline
Word Rep. &           0.500            &  0.689 \\
Joint Word-Category Rep. &     0.526 \small(+5.0\%)       &   0.711 \small(+3.1\%)\\
Joint Word-Category Rep. + $\mathcal{L}_{CM}$&    \bf  0.540  \small(+2.9\%) & \bf 0.720 \small(+1.3\%)\\

\hline
\end{tabular}
\caption{\small Ablation analysis results.}
\label{tab:ablation}
\vspace{-5mm}
\end{table}

\textbf{\textit{ Summary: }} Our experimental results show the robust performance of DeepCAT compared to state-of-the-art models. For {\em minority} classes, {\em tail}, and {\em torso} queries, we observed 10\%, 7\%, and 5.3\% relative improvements, respectively. We also report the performance on the last layer (leaf nodes) of product taxonomy consisting of 4115 categories. The results show that DeepCAT achieves (6.2\%, 4\%, 3\%, and 3\%) increase on \textit{F1@1, F1@3, F1@5,} and \textit{MAP@5}, respectively. In ablation analysis, we show that the improvements come from all three components of DeepCAT. The joint word-category representation improves the query representation by 5\%, and the loss function can further improve it by 2.9\%. 

\section{Conclusions}
We introduced a deep learning model, DeepCAT, for query understanding in e-commerce search. DeepCAT contains a new joint word-category representation component in which category representations are learned using word-category co-occurrences.
Then, we proposed a novel loss function utilizing category representations to model category-category co-occurrences. Our comprehensive experiments showed that using category representation significantly improved the results, particularly on {minority} classes and {\em tail} queries. DeepCAT achieved a 10\% improvement on {\em minority} classes and a 7.1\% increase on {\em tail} queries over a state-of-the-art label embedding model. 

\subsubsection*{\bf Acknowledgements}
We gratefully acknowledge the financial and computing support from The Home Depot Search \& NLP team.

\balance
\bibliographystyle{abbrv}
\bibliography{Reference}

\end{document}